\documentclass[preprint,aps,12pt,preprintnumbers,eqsecnum,nofootinbib,superscriptaddress]{revtex4}

\usepackage{amssymb,amsmath}
\usepackage[mathscr]{euscript}
\usepackage{mathrsfs} 
\usepackage[usenames,dvipsnames]{color}
\usepackage{float, subfigure}
\usepackage{setspace}
\usepackage{graphicx}

\usepackage{simplewick} 
\usepackage{epstopdf} 

\DeclareMathOperator{\tr}{tr}
\everymath{\displaystyle}
\renewcommand{\cal}{\mathcal}
\newcommand\relphantom[1]{\mathrel{\phantom{#1}}}

\begin{document}
%
\title{\vspace*{0.5in} 
Curved Backgrounds in Emergent Gravity
\vskip 0.1in}
\author{Shikha Chaurasia}\email[]{sschaurasia@email.wm.edu}\affiliation{High Energy Theory Group, Department of Physics,
College of William and Mary, Williamsburg, VA 23187-8795, USA}
\author{Joshua Erlich}\email[]{jxerli@wm.edu}\affiliation{High Energy Theory Group, Department of Physics,
College of William and Mary, Williamsburg, VA 23187-8795, USA}
\author{Yiyu Zhou}\email[]{yzhou11@email.wm.edu}\affiliation{High Energy Theory Group, Department of Physics,
College of William and Mary, Williamsburg, VA 23187-8795, USA}
\date{October 31, 2017}
%
%

%
%

\begin{abstract}
Field theories that are generally covariant but nongravitational at tree level typically give rise to an emergent gravitational interaction whose strength depends on a physical regulator.  We consider emergent gravity models in which scalar fields assume the role of clock and rulers, addressing the problem of time in quantum gravity. We discuss the possibility of nontrivial dynamics for clock and ruler fields, and describe some of the consequences of those dynamics for the emergent gravitational theory.
\end{abstract}
\maketitle

\section{Introduction}

The possibility that gravitation emerges from other interactions provides a promising paradigm for addressing the difficult
conceptual questions that confront quantum gravity. These questions include the problem of time, namely that coordinate invariance implies a vanishing Hamiltonian and the consequent absence of dynamics of quantum states \cite{DeWitt:1967yk}; the question of predictivity in a theory with nonrenormalizable interactions such as gravitation; the question of what observables are physical in a diffeomorphism-invariant theory; and questions related to the vacuum, including why the Minkowski spacetime and its signature should be preferred to other spacetimes in a quantum theory in which spacetime geometries are integrated over. 

The possibility of emergent long-range interactions in quantum field theory has been recognized for half a century.\footnote{In a classical context, Michael Faraday suspected a relationship between electromagnetism and gravitation, and in the 1840s searched experimentally for such an identification. He was unsuccessful
\cite{Faraday}.}  
 Bjorken argued that four-fermi models with current-current interactions can give rise to emergent gauge interactions \cite{Bjorken:1963vg}, and Eguchi later argued that the  composite gauge field in such theories may render those theories renormalizable despite the presence of fundamental four-fermi interactions \cite{Eguchi:1976iz}.
It did not take long for the idea of emergent interactions to be extended to gravitation, in a wonderfully short note by Sakharov \cite{Sakharov:1967pk}. Sakharov pointed out that the regularized effective action for the spacetime metric  generically contains the Einstein-Hilbert term even if no such term is present at tree level, as long as general covariance is maintained by the regulator in the theory. This suggests that the dynamics of spacetime might emerge as an artifact of regulator-scale physics even if there is no such dynamics prior to quantization.\footnote{Sakharov had in mind that the spacetime metric was to be treated classically, in which case the induced gravity is semiclassical, with the vacuum expectation value of the energy-momentum tensor $T_{\mu\nu}$ being related to the spacetime metric by Einstein's equations (with cosmological constant, plus regulator-suppressed corrections).}  

Perhaps the most compelling argument for emergent gravity is its ubiquity: all that is needed is a generally covariant description of the interactions of a field theory and a covariant regulator that resolves infinities in perturbation theory, both of which are likely to be required of quantum gravity, in any case. Much work has been done  in an attempt to turn Sakharov's observation into a compelling description of quantum gravity \cite{Akama:1978pg,Akama:1977hr,Amati:1981rf,Amati:1981tu}, but certain difficulties remain. More recently, alternative paradigms that also appear to lead to emergent gravitational interactions have gained favor, such as the AdS/CFT correspondence \cite{Maldacena:1997re}, entropic gravity \cite{Verlinde:2010hp}, and emergent spacetime via networks of entangled states \cite{Maldacena:2013xja,Cao:2016mst}. However,
the present work concerns the old-fashioned approach to the subject.

The problem of nonrenormalizability of the gravitational interaction persists in emergent gravity, unless the quantum theory is asymptotically safe by virtue of an ultraviolet fixed point \cite{Weinberg}. However, with the presumption of a physical regulator, the lack of predictivity of the theory is augmented by the more fundamental ontological question of what is to be demanded of the theory at short distances. 
Regulators in quantum field theory have the habit of violating some cherished principle or another, such as unitarity or  boundedness of the Hamiltonian from below. In the present work we are agnostic about the physical regulator and its consequences for the interpretation of the theory at short-distances, and we require only that the theory provide a definite rule for calculating correlation functions of appropriate observables at all physical scales. For the purpose of illustration we will use dimensional regularization, fixing the spacetime dimension $D$ by holding $\epsilon=D-4$ small but fixed.

The problem of time demands that physical degrees of freedom playing the roles of clock and rulers  be identified in any generally covariant theory. This allows dynamics to be interpreted in terms of correlations, or entanglement \cite{Page:1983uc},  between physical degrees of freedom and the clock and rulers.  For example, certain scalar fields $X^J(x^\mu)$ can play the role of the physical clock and rulers 
by a gauge-fixing condition analogous to the static-gauge condition in string theory, under the presumption that field configurations dominating the functional integral can be put into that gauge. Here $x^\mu$ are the spacetime parameters integrated over in the action, and  the indices $J$ and $\mu$ both take values in $\left\{0,...,D-1\right\}$.  The gauge choice is $X^J= c\,x^\mu\,\delta_\mu^J$ for some constant $c$ that will be specified for convenience later. In the models considered here, this choice for the fields $X^J$  satisfy the classical equations of motion with all other fields sitting at the minimum of the potential, and there is a natural perturbative expansion about this classical background.

In this note we generalize a particular toy model of emergent gravity that was recently studied in Ref.~\cite{Carone:2016tup}. The model contains only scalar fields, and  $D$ of the fields play the role of clock and rulers in $D$ dimensions. The model was shown to include a massless composite graviton in its spectrum which couples at leading order  to the energy-momentum tensor of the physical (non-gauge-fixed) fields as in Einstein gravity. The model is generally covariant from the outset, has a vanishing energy-momentum tensor (including the contributions of the clock and ruler fields), and thereby evades the Weinberg-Witten no-go theorem which prohibits the existence of massless spin-2 particles in a broad class of Lorentz-invariant theories \cite{Weinberg:1980kq}. Here we generalize the theory to the case in which the clock and ruler fields have a nontrivial field-space metric, and we demonstrate that, at leading order
in a perturbative expansion, scattering is as in Einstein gravity in a spacetime background identified with the field-space metric.

\section{Emergent Gravity with Curved Backgrounds}

The theory that we study includes $N$ scalar fields $\phi^a$, $a\in\left\{1,\dots,N\right\}$, in addition to the $D$ scalar fields $X^J$ that play the role of clocks and rulers. We assume the potential depends only on $\phi^a$ but not the clock and ruler fields. The theory is defined so as to be diffeomorphism invariant, and at the classical level the theory is independent of any geometric structure imposed on the spacetime other than differentiability. In particular, the action is independent of spacetime metric on the coordinates $x^\mu$, and correspondingly the theory has an identically vanishing energy-momentum tensor.  The action for the theory is, 
\begin{equation}
S=\int \mathrm{d} ^Dx\ \left(\frac{\tfrac D2-1}{V(\phi^a)} \right)^{\frac{D}{2}-1}
\sqrt {\left|\det \left(\sum_{a=1}^N \partial_\mu\phi^a \,\partial_\nu\phi^a 
+\sum_{I,J=0}^{D-1}\partial_\mu X^I \,\partial_\nu X^J\, G_{IJ}(X^K)\right)\right|}.
\label{eq:S}\end{equation}
Aside from the dependence of the action on a potential $V(\phi^a)$, this theory is in the class of induced gravity theories based on Dirac's membrane action, as analyzed recently in Ref.~\cite{Akama:2013tua}. 

The theory described by Eq.~(\ref{eq:S}) is nonlinear, but it is reminiscent of the Nambu-Goto action for the string and we can motivate it  by introducing an auxiliary  spacetime metric which is fixed by a constraint of vanishing energy-momentum tensor. The Polyakov-like description of the theory is given by the action,
\begin{equation} \label{eq:S1}
S=\int \mathrm{d} ^Dx\,\sqrt{|g|}\left[\frac{1}{2}g^{\mu\nu}\left(\sum_{a=1}^N\partial_\mu\phi^a\,\partial_\nu\phi^a+\sum_{I,J=0}^{D-1}\partial_\mu X^I\partial_\nu X^JG_{IJ}(X^K)\right)
-V(\phi^a)\right].
\end{equation}
The quantum theory is defined by functional integral quantization over the scalar fields and $g_{\mu\nu}(x)$, subject to the constraint $T_{\mu\nu}=0$.  The constraint can be thought of as arising from integrating out the auxiliary-field metric $g_{\mu\nu}$, although in that case a Jacobian functional determinant would also arise from the functional integration, which would appear to modify the theory nonperturbatively. (An analogy to this in the context of a model of emergent gauge interactions was pointed out in Ref.~\cite{Suzuki:2016aqj}.)
The partition function for the theory is,
\begin{equation}
Z = \int_{T_{\mu\nu}=0} \cal Dg_{\mu\nu}\, \cal D\phi^a\, \cal D X^I\, e^{i S(\phi^a, X^I, g_{\mu\nu})}, 
\end{equation}
where the symmetric energy-momentum tensor is defined in the usual way,
\begin{eqnarray}
T_{\mu\nu}(x)&=&\frac{2}{\sqrt{|g|}}\frac{\delta S}{\delta g^{\mu\nu}(x)} \label{eq:Tmn} \\
&=&\sum_{a=1}^N\partial_\mu \phi^a \partial_\nu\phi^a +\sum_{I,J=0}^{D-1}\partial_\mu X^I\partial_\nu X^JG_{IJ}-g_{\mu\nu} {\cal L},\label{eq:Tmn2}\end{eqnarray}
where the Lagrangian ${\cal L}$ is defined by the action in Eq.~(\ref{eq:S1}), $S\equiv\int \mathrm{d}^Dx\,\sqrt{|g|}{\cal L}$.
Eq.~(\ref{eq:Tmn2}) is solved by 
\begin{equation}
g_{\mu\nu}=\frac{D/2-1}{V(\phi^a)}\left(\sum_{a=1}^N\partial_\mu \phi^a\partial_\nu\phi^a+
\sum_{I,J=0}^{D-1}\partial_\mu X^I\partial_\nu X^JG_{IJ}\right),
\label{eq:gmn}\end{equation}
which together with Eq.~(\ref{eq:S1}) gives the Dirac brane-like action Eq.~(\ref{eq:S}). 
We assume that the potential $V(\phi)$ has the form $V(\phi)=V_0+\Delta V(\phi)$, with the minimum of the potential $V_0$ much larger than any other scales in the theory with the possible exception of a scale associated with the physical regulator. For simplicity we also assume in our analysis that the field-space metric $G_{IJ}(X^K)\equiv\eta_{IJ}+\widetilde{H}_{IJ}(X^K)$, with Minkowski (mostly-minus) metric $\eta_{IJ}$, admits a perturbative expansion in $\widetilde{H}_{IJ}$ and its derivatives.

The theory described by Eq.~(\ref{eq:S}) is invariant under coordinate reparametrizations, $X^I(x)\rightarrow X^I(x^\prime(x))$ and $\phi^a(x)\rightarrow \phi^a(x^\prime(x))$; and under  field redefinitions the field-space metric $G_{IJ}$ transforms like a metric: If $X^I(x)$ is replaced with $X^{\prime\,I}(X^J(x))$, \begin{eqnarray}
\partial_\mu X^I \partial_\nu X^J G_{IJ}(X) &\rightarrow& \partial_\mu X^{\prime\,I}\partial_\nu X^{\prime\,J} G_{IJ}(X^\prime(X))
\nonumber \\
&&=\partial_\mu X^K \partial_\nu X^L  \frac{\partial X^K}{\partial X^{\prime\,I}} \frac{\partial X^L}{\partial X^{\prime\,J}} G_{IJ}(X^\prime(X)) \ \\
&&=\partial_\mu X^I \partial_\nu X^J G^\prime_{IJ}(X), \nonumber \end{eqnarray}
where \begin{equation}
G^\prime_{IJ}(X)=\frac{\partial X^K}{\partial X^{\prime\,I}} \frac{\partial X^L}{\partial X^{\prime\,J}} G_{IJ}(X^\prime(X)).
\label{eq:fieldredef}\end{equation}
Note that a field redefinition cannot take a curved-space $G_{IJ}$ to a flat-space one, so the theory with generic field-space metric is genuinely inequivalent to the flat-field-space version of the theory studied previously.

In order to provide physical meaning to the spacetime background in which dynamics take place, we identify $X^I$ with the corresponding spacetime coordinates (up to a constant factor), analogous to a static gauge condition in string theory:
\begin{equation}
\label{eq:sgc}
X^I = \sqrt{\frac{V_0}{D/2- 1}} x^\mu \delta_\mu ^I, \quad I = 0, \ldots, D-1,
\end{equation}
Then the field $X^0$ can be interpreted as an internal clock \cite{Page:1983uc}, while the fields $X^i, i = 1, \ldots, D-1$ are interpreted as rulers. In this case the Fadeev-Popov determinant is 
\begin{equation}\label{eq:FPD}
\begin{split}
\det\left(\frac{\delta X^{I , \alpha} \left( y \right)}{\delta \alpha ^{\mu} \left( y ' \right)}\right)
& =\det\left(\sqrt{\frac{V_0}{\frac{D}{2} - 1}} \frac{\delta \left( y^{\mu} + \alpha ^{\mu} \left( y \right) \right)}{\delta \alpha ^{\mu} \left( y' \right)} \delta _{\mu}^I\right) \\
& = \det\left(\sqrt{\frac{V_0}{\frac{D}{2} - 1}} \delta _{\mu}^I \delta ^{\left( D \right)} \left( y - y' \right)\right) , \\
\end{split}
\end{equation}
which is trivial and consequently there are no Fadeev-Popov ghosts  resulting from gauge fixing $X^I$. 

The classical equations of motion for $\phi ^a$ and $X^I$ following from the action Eq.~(\ref{eq:S1}) are
\begin{equation}
\label{eq:phieom}
\frac{1}{\sqrt{- g}} \partial _{\mu} \left( \sqrt{-g} g^{\mu \nu} \partial _{\nu} \phi ^a \right) = - \frac{\partial V}{\partial \phi ^a},
\end{equation}
\begin{equation}
\label{eq:xeom} \partial _{\mu} \left( \sqrt{-g} g^{\mu \nu} G_{I J} \partial _{\nu} X ^J \right) = \frac{1}{2} \sqrt{-g} g^{\mu \nu} \partial _{\mu} X^J \partial _{\nu} X^K \frac{\partial}{\partial X^I} G_{J K}.
\end{equation}
If we set $\phi ^a  = \phi^a _{\min}$ where $\phi^a_{\min}$ minimizes $V$ such that $ V(\phi^a_{\min})=V_0$, then the equation of motion for $\phi ^a$ is trivially satisfied. Meanwhile, with the gauge-fixed background $X^I$ as in Eq.~(\ref{eq:sgc}), the spacetime metric at $\phi^a = \phi^a_{\min}$ is
\begin{equation}
g _{\mu \nu}= \frac{\frac{D}{2} - 1}{V_0} \frac{\left( \sqrt{V_0} \delta _{\mu}^I \right) \left( \sqrt{V_0} \delta _{\nu}^J \right) G_{I J}}{\frac{D}{2} - 1} = G_{\mu \nu}( x^I),
\end{equation}
so the spacetime background in which the fields $\phi^a$ propagate is now identified with the field-space metric for the clock and ruler fields. Furthermore, the equations of motion for the clock and ruler fields, Eq.~(\ref{eq:xeom}), are also satisfied by the static gauge condition, as is readily checked using the identity,
\begin{equation}
\frac{1}{\sqrt{|g|}} \frac{\partial \sqrt{|g|}}{\partial x^\alpha} = \frac{1}{2} g^{\mu \nu} \frac{\partial g_{\mu \nu}}{\partial x^\alpha}.\end{equation}
Hence, the static-gauge configuration with fields $\phi^a$ uniform at the minimum of the potential, and with $g_{\mu\nu}(x)=G_{\mu\nu}(x)$, solve the equations of motion and provide a classical background about which the dynamics for the fields $\phi^a$ can now be analyzed.

We now show that the background $G_{IJ}$ modifies the emergent gravitational interaction by coupling to the matter fields as in Einstein gravity, at linear order in the expansion about the Minkowski metric. Thus we  write the background $G_{I J}$ as,
\begin{equation} 
G_{\mu\nu} = g_{\mu\nu}^{ \left(B \right)} = \eta_{\mu\nu} + \widetilde{H}_{\mu\nu},
\end{equation}
where $\widetilde H_{\mu\nu}$ determines the background spacetime but is assumed to be small compared to $\eta_{\mu\nu}$. Consequently the gauge-fixed action takes the form,
\begin{equation}
\label{eq:gaugeact}
S = \int  \mathrm{d}^D{x} \frac{V_0}{D/2-1}\left(\frac{V_0}{V_0 + \Delta V(\phi^a)}\right)^{D/2-1}\sqrt{\left|\text{det}\left(\eta_{\mu\nu}+\widetilde H_{\mu\nu} + \widetilde h_{\mu\nu}\right)\right|},
\end{equation}
where 
\begin{equation}
\label{eq:h}
 \widetilde h_{\mu\nu} \equiv \frac{D/2-1}{V_0} \left(\sum_{a=1}^N \partial_\mu \phi^a \partial_\nu \phi^a\right),
 \end{equation}
and  $g_{\mu \nu}$ depends on the field configuration via,
\begin{equation} 
g_{\mu\nu} = \frac{V_0}{V(\phi)} \left(\eta_{\mu\nu} + \widetilde H_{\mu\nu} + \widetilde h_{\mu\nu}\right). 
\end{equation}

In order to analyze the theory perturbatively, we expand Eq.~(\ref{eq:gaugeact}) in powers of $1/V_0$ and $\widetilde H$. We take $\widetilde{h}_{\mu\nu}$ and $\widetilde H_{\mu\nu}$ to be of the same order. We also assume for simplicity that $N$, the number of fields $\phi^a$, is large, and keep only leading terms in a $1/N$ expansion. Expanding the determinant via the identity $\det M = \exp{\left( \tr \ln M \right)}$, the action can be written as 
\begin{equation}
\begin{split}
S &= \int  \mathrm{d}^D{x} \frac{V_0}{D/2-1}\left( 1+\frac{\Delta V(\phi^a)}{V_0}\right)^{1-D/2} \left[1 + \frac{1}{2}\left(\widetilde h+ \widetilde H \right) \right.
\\
& \relphantom{= \int  \mathrm{d}^D{x}} \left. - \frac{1}{4}\left(\widetilde h_{\mu\nu} + \widetilde H_{\mu\nu}\right)\left(\widetilde h^{\mu\nu} + \widetilde H^{\mu\nu}\right)+\frac{1}{8}\left(\widetilde h + \widetilde H \right)^2+\cdots \right] \\
&= \int  \mathrm{d}^D{x} \left(\frac{V_0}{D/2-1} - \Delta V(\phi^a) + \frac{D}{4} \frac{\left(\Delta V(\phi^a) \right)^2}{V_0} +\cdots\right) \times \left[1+\frac{1}{2}\left(\widetilde h+ \widetilde H\right) \right. \\
& \relphantom{= \int  \mathrm{d}^D{x}} \left. - \frac{1}{4}\left(\widetilde h_{\mu\nu} \widetilde h^{\mu\nu} +\widetilde H_{\mu\nu} \widetilde H^{\mu\nu}+ 2\widetilde h_{\mu\nu} \widetilde H^{\mu\nu}\right) + \frac{1}{8}\left(\widetilde h^2 + \widetilde H^2 + 2 \widetilde h\widetilde H\right)+\cdots\right],
\end{split}
\end{equation}
where index contractions are done with the Minkowski metric and $\widetilde h = \eta_{\mu\nu} \widetilde h^{\mu\nu}$ (likewise $ \widetilde H = \eta_{\mu\nu} \widetilde H^{\mu\nu}$). Keeping terms up to first order in $\widetilde H$ and $1/V_0$, and using Eq.~(\ref{eq:h}), we arrive at the action
\begin{equation}
\begin{split}
 S = \int  \mathrm{d}^D{x} & \left\{ \frac{V_0}{D/2-1} + \frac{1}{2}\sum_{a=1}^N \partial_\mu\phi^a\partial^\mu\phi^a - \Delta V(\phi^a) + \frac{1}{2} \frac{V_0}{D/2 - 1} \widetilde{H} \right. \\
& \left. {} - \frac{D/2-1}{4V_0}\left[ \sum_{a=1}^N \partial_\mu\phi^a\partial_\nu\phi^a \sum_{b=1}^N \partial^\mu\phi^b\partial^\nu\phi^b - \frac{1}{2}\left(\sum_{a=1}^N \partial_\mu\phi^a\partial^\mu\phi^a\right)^2\right] \right. \\
& \left. {} -\frac{D/2-1}{2}\frac{\Delta V(\phi^a)}{V_0}\sum_{a=1}^N \partial_\mu\phi^a\partial^\mu\phi^a + \frac{D}{4}\frac{\left(\Delta V(\phi^a) \right)^2}{V_0} \right. \\
& \left. {} - \frac{1}{2}\widetilde H^{\mu\nu}\sum_{a=1}^N \partial_\mu\phi^a\partial_\nu\phi^a + \frac{1}{4}\eta_{\mu\nu}\widetilde H^{\mu\nu} \sum_{a=1}^N \partial_\alpha \phi^a\partial^\alpha \phi^a - \frac{1}{2}\Delta V(\phi^a) \eta_{\mu\nu} \widetilde H^{\mu\nu} \right. \\
& \left. {} + \mathcal{O}\left(\widetilde{H}^2, \, \frac{1}{V_0^2}\right)  \right\}.
\end{split} \label{eq:expact}
\end{equation}
The first three lines in Eq.~(\ref{eq:expact}) are equivalent to the action analyzed in Ref.~\cite{Carone:2016tup} up to the addition of a $\phi$-independent contribution to the action.

The interactions between $\phi^a$ and $\widetilde H^{\mu\nu}$ are new, and will shortly be shown to give rise to scattering off of the background spacetime in accordance with general relativity. For a free theory with O$(N)$-symmetric potential
\begin{equation}
\Delta V(\phi^a) = \sum_{a=1}^N \frac{m^2}{2}\phi^a\phi^a,
\end{equation} 
the first line of Eq.~(\ref{eq:expact}) contains the free part of the action. The energy-momentum tensor for free fields $\phi^a$ is 
\begin{equation}
\label{eq:T}
\cal T_{\mu\nu} = \sum_{a=1}^N \left[\partial_\mu\phi^a\partial_\nu\phi^a - \eta_{\mu\nu}\left(\frac{1}{2}\partial^\alpha\phi^a\partial_\alpha\phi^a-\frac{1}{2}m^2\phi^a\phi^a\right)\right],
\end{equation}
and the interacting terms excluding $\widetilde H^{\mu\nu}$ can be written,
\begin{equation}
\mathcal{L} _{\widetilde{h}} = - \frac{1}{4 V_0} \mathcal{T} _{\mu \nu} \mathcal{T} _{\alpha \beta} \left( \left( \frac{D}{2} - 1 \right) \eta ^{\nu \alpha} \eta ^{\mu \beta} - \frac{1}{2} \eta ^{\mu \nu} \eta ^{\alpha \beta} \right).
\end{equation}
In Ref.~\cite{Carone:2016tup}, it was shown that these interactions give rise to a massless spin-two graviton state that mediates the gravitational interaction in two-into-two scattering of $\phi$ bosons. 

\begin{figure}[H]
\centering
\includegraphics[scale=0.8]{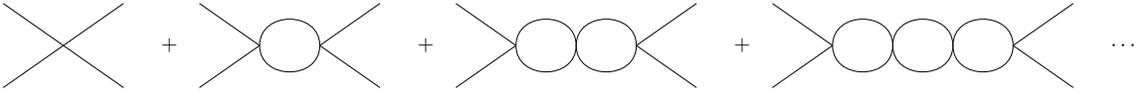}\\
\caption{Leading large-$N$ diagrams that give rise to the emergent gravitational interaction.}
\label{fig:BC}
\end{figure}

\begin{figure}[H]
\centering
\hfil
\subfigure[The graviton pole that emerges from two-into-two scattering of $\phi$ bosons.]{
\includegraphics[scale=1.7]{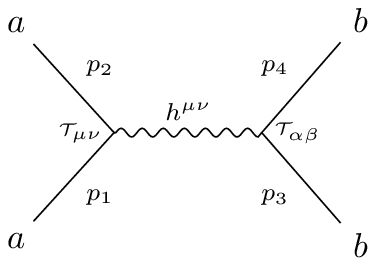}
\label{fig:MMA}
}
\hfil
\subfigure[Scattering of $\phi$ bosons off of $\widetilde{H}^{\mu \nu}$.]{
\includegraphics[scale=1.7]{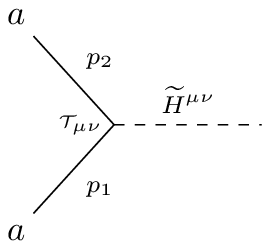}
\label{fig:MBA}
}
\hfil
\caption{Feynman diagrams for our current theory.}
\label{fig:PT}
\end{figure}
The diagrams in Fig.~\ref{fig:BC} are responsible for the emergent gravitational interaction, which can be equivalently described by exchange of a composite graviton as in Fig.~\ref{fig:MMA}. Hence, the emergent gravity persists in this model, at least at the perturbative level to which we are working.

At the same order in perturbation theory, we can interpret the interactions with the background metric $\widetilde H^{\mu\nu}$  as arising from a background source. Notice the contribution from $\widetilde H^{\mu\nu}$ in the last line of Eq.~(\ref{eq:expact}) takes the form
\begin{equation}
\label{eq:HT}
\mathcal{L} _{\widetilde{H}} =  -\frac{1}{2}\widetilde H^{\mu\nu} \sum_{a=1}^N \left[ \partial_\mu\phi^a \partial_\nu\phi^a -\eta_{\mu\nu}\left(\frac{1}{2}\partial^\alpha\phi^a\partial_\alpha\phi^a - \frac{1}{2}m^2\phi^a\phi^a\right)\right]  = -\frac{1}{2} \widetilde H^{\mu\nu} \cal T_{\mu\nu},
 \end{equation}
 which confirms the agreement of the theory with the linearized coupling of matter to the background metric in general relativity, and results in the interactions shown in Fig.~\ref{fig:MBA}.
From Eq.~(\ref{eq:HT}), we can read off the momentum space Feynman rule for interactions involving $\widetilde H^{\mu\nu}$, with $p_1$ ingoing and $p_2$, $q$ outgoing:
\begin{equation}
\left(\widetilde H-{\cal T}\right)\ {\rm vertex}=-\frac{i}{2}E_{\mu\nu} \left(p_1, p_2\right) \widetilde H^{\mu\nu} \left(q\right) \delta^D\left(p_1-p_2-q\right)
\end{equation}
for inwardly (outwardly) directed external momenta $p_1$ ($p_2$), and where 
\begin{equation}
E_{\mu\nu}\left(p_1, p_2\right) \equiv \left(p_1^\mu p_2^\nu + p_1^\nu p_2^\mu\right)+\eta^{\mu\nu} \left(-p_1 \cdot p_2+m^2\right)
\end{equation}
is determined by Eq.~(\ref{eq:T}), summing over the ways in which the fields can annihilate (or create) incoming (or outgoing) scalar bosons. The interactions involving $\widetilde H^{\mu\nu}$ don't contribute to scattering but instead create an instability in $G_{\mu\nu}$, rendering $\mathcal{T}_{\mu\nu}\neq0$. Hence there is a background field (call this ${\cal T^{ \left(B \right)}}_{\mu\nu}$) that appears as a source for $\widetilde H^{\mu\nu}$ in the Einstein-Hilbert action.
 
We note that interactions at higher-order in $1/V_0$ can contribute at the same order as the diagrams that we have considered if they include tadpoles which are also proportional to $V_0$.  However, as in Ref.~\cite{Carone:2016tup}, we can add a counterterm $c_2$ to $V_0$ which cancels tadpoles from insertions of $m^2\contraction{}{\phi ^a}{}{\phi ^a} \phi^a  \phi^a$ in interactions, and we can shift the gauge by a parameter $c_1$ in order to cancel tadpoles from insertions of $\contraction{\partial _{\mu}}{\phi ^a}{\partial _{\nu}}{\phi ^a} \partial_{\mu} \phi ^a \partial_\nu \phi ^a$ in interactions:

\begin{equation}
\label{eq:rescale}
\begin{split}
X^I &= x^I\sqrt{\frac{V_0}{D/2-1}-c_1},
\\ \Delta V &= \frac{1}{2}m^2\phi^a\phi^a - c_2,
\end{split}
\end{equation}
 There are no other tadpoles in this theory, so all relevant diagrams have been accounted for at leading order in $1/N$ and $1/V_0$. All additional diagrams from couplings of higher order in $1/V_0$ are consistently neglected at leading order.

The linearized coupling of the composite field $h _{\mu \nu}$ to matter is given by
\begin{equation}
 \cal L_{hT} = -\frac{1}{2} h^{\mu\nu} \cal T_{\mu\nu},
\end{equation} 
where $h^{\mu\nu}$ is the composite operator representing the fluctuation about the Minkowski metric,
\begin{equation}
\label{eq:tensorstructure}
\begin{gathered}
 h^{\mu\nu} = \frac{1}{V_0} P^{\mu\nu}\,_{\lambda\kappa} \cal T^{\lambda\kappa} + \cal O\left(1/V_0^2\right) = \frac{1}{V_0}\sum_{a=1}^N\left[\left(D/2-1\right)\partial^\mu\phi^a\partial^\nu\phi^a - \frac{1}{2}\eta^{\mu\nu} m^2\phi^a\phi^a\right], \\
P^{\mu\nu}\,_{\lambda\kappa}  \equiv \frac{1}{2}\left[ \left(D/2-1\right) \left(\delta^\mu_\lambda\delta^\nu_\kappa+\delta^\mu_\kappa \delta^\nu_\lambda\right) - \eta^{\mu\nu}\eta_{\lambda\kappa} \right]. \\
\end{gathered}
\end{equation}
Now that there is a source creating a background in which $T_{\mu\nu}$ fluctuates, we find that 
\begin{equation}
\label{eq:hHT}
\cal L_{int} = -\frac{1}{2}\left(h^{\mu\nu}+\widetilde H^{\mu\nu} \right)\cal T_{\mu\nu}
\end{equation}
at the linearized level.

Thus we have interactions in which the matter fields can scatter off themselves, corresponding to the exchange of a massless composite graviton $h^{\mu\nu}$, or they can scatter off the background spacetime defined by $\widetilde H^{\mu\nu}$ . We can interpret the scattering off of the background spacetime as due to the existence of a background energy-momentum tensor. Here we can draw an analogy to electromagnetism. Consider a scenario in which there is a current creating a background electromagnetic field; then incoming charged particles feel the effects of the field as they scatter off of one another.  But we can recast this scenario into an equivalent one in which the incoming charged particles scatter off the current which generates the background electromagnetic field, thereby rendering the source dynamical.

\begin{figure}[H]
\centering
\includegraphics[scale=1.7]{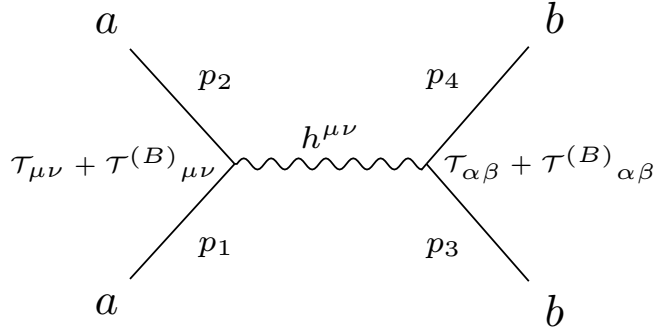}\\
\caption{In this redefined theory, the matter field $\mathcal{T}_{\mu \nu}$ can scatter off of itself and the background field ${\mathcal{T}^{\left( B \right)}} _{\mu \nu}$. The scattering of ${\cal{T}^{ \left(B \right)}}_{\mu \nu}$ off of itself is unphysical, thus should not be considered.}
\label{fig:NT}
\end{figure}

Likewise we can consider a process in which the scalar bosons scatter off of one another and off of the source that generates $\widetilde H^{\mu\nu}$, so that the graviton couples to the matter and background source, as shown in Fig.~\ref{fig:NT}. As a result the interacting Lagrangian reads
\begin{equation}
\label{eq:htnew}
{\cal L}_{int}' = -\frac{1}{2}  h^{\mu\nu}\left({\cal T}_{\mu\nu} + {\cal T^{\left(B\right)}}_{\mu\nu} \left(x\right)\right).
\end{equation}

We can extract ${\cal T^{\left(B\right)}}_{\mu\nu}$ from the linearized equation of motion for $\widetilde H^{\mu\nu}$:
\begin{equation}
\label{eq:LEE}
D_{\mu\nu\lambda\kappa}\widetilde H^{\lambda\kappa}(x) = -{\cal T^{ \left(B \right)}}_{\mu\nu}(x).
\end{equation}
Here $D$ is the linearized equation of motion operator describing the dynamics of the composite graviton,
\begin{equation}
\label{eq:origD}
\begin{split}
D_{\mu\nu\lambda\kappa}\widetilde H^{\lambda\kappa}(x) & = \frac{D-2}{M_P^{2-D}}\left(
\square \,\widetilde H^{\mu\nu} + \partial^{\mu} \partial^{\nu} \widetilde H - \eta^{\mu\nu}\,\square\, \widetilde H \right. \\
& \relphantom{=\frac{D-2}{M_P^{2-D}}(} \left. {} + \eta^{\mu\nu} \partial_\lambda \partial_\kappa \widetilde H^{\lambda\kappa} - \eta^{\nu\lambda} \partial_\lambda \partial_\kappa \widetilde H^{\mu\kappa} - \eta^{\nu\lambda}\partial ^\mu \partial^\kappa \widetilde H_{\lambda\kappa}\right),
\end{split}
\end{equation}
where $M_P$ characterizes the strength of the interaction. It is the reduced Planck mass deduced by comparing the scattering amplitude due to the effective 1-graviton exchange with general relativity, and was calculated in Ref.~\cite{Carone:2016tup} to have the value\footnote{We thank Chris Carone and Diana Vaman for correcting a minus sign in this expression from the first version of Ref.~\cite{Carone:2016tup}.},
\begin{equation}
M_P = m\left[ \frac{N\Gamma \left(1-D/2\right)}{6\left(4\pi\right)^{D/2}}\right]^{1/\left(D-2\right)}.
\end{equation}
If we instead want to recover the background spacetime from the background energy-momentum tensor, we need to invert the equation of motion operator, which requires fixing the coordinate ambiguity. For example, by a field redefinition of the fields $X^I$ as in Eq.~(\ref{eq:fieldredef}), we can choose the background $\widetilde H^{\mu\nu}$ to be in the de Donder gauge, $\partial ^{\nu} \widetilde H_{\mu \nu} = \tfrac{1}{2}\partial _{\mu} \widetilde H$.  The linearized Einstein equation in de Donder gauge is,
\begin{equation}
D_{\mu\nu\lambda\kappa}\widetilde H^{\lambda\kappa}(x) = \frac{D-2}{M_P^{2-D}}\left( \square \widetilde{H} ^{\mu \nu} - \frac{1}{2} \eta ^{\mu \nu} \square \widetilde{H}\right).
\end{equation}
Since this expression is invertible, we can calculate schematically,
\begin{equation}
\label{eq:Dinvor}
\widetilde H^{\mu\nu} = - \left( D^{-1}\right)^{\mu\nu\lambda\kappa}{\cal T^{ \left(B \right)}}_{\lambda\kappa}.
\end{equation}
Upon a Fourier transformation to momentum space, 
\begin{equation}
\label{eq:Dinv}
\widetilde{H}^{\mu\nu}(q) =-  \left(D^{-1} \right)^{\mu\nu\lambda\kappa}(q) {E^{ \left(B \right)}}_{\lambda\kappa} \left(p_3, p_4 \right),
\end{equation}
for incoming (outgoing) momenta $p_3$ ($p_4, q=p_3-p_4$).

We can compare in more detail the theories defined by Lagrangians Eq.~(\ref{eq:hHT}) and Eq.~(\ref{eq:htnew}). Seeing that both $\cal L_{int}$ and $\cal L'_{int}$ contain a factor of $h^{\mu\nu}\cal T_{\mu\nu}$, the scattering amplitude of scalar particles off one one another remains the same at this order, so we only need to consider interactions involving the source and the background it creates. The scattering amplitude for Fig.~\ref{fig:MBA} is
\begin{equation}
\mathcal{M}_{MB} = - \frac{i}{2} E_{\mu \nu} \left( p_1, p_2 \right) \widetilde{H}^{\mu \nu}( q),
\end{equation}
while the scattering amplitude for  Fig.~\ref{fig:NT} is
\begin{equation}
\begin{split}
\mathcal{M}'_{MB} & = - \frac{i}{2} E_{\mu \nu} \left( p_1 , p_2 \right) (-i) \left( D^{- 1} \right) ^{\mu \nu \alpha \beta}( q) 2 \times \left( - \frac{i}{2} {E^{\left( B \right)}}_{\alpha \beta} \left( p_3 , p_4 \right) \right) \\
& = - \frac{i}{2} E_{\mu \nu} \left( p_1 , p_2 \right) 2 \times \left( \frac{1}{2} {\widetilde{H}}^{\mu \nu}(q) \right) \\
& = - \frac{i}{2} E_{\mu \nu} \left( p_1 , p_2 \right) {\widetilde{H}}^{\mu \nu} (q), \\
\end{split}
\end{equation}
Evidently $\cal M_{MB} = \cal M_{MB}'$; thus we can infer that the amplitude of the matter fields scattering off of the background source is the same as if it was scattering off of the background metric. Indeed then ${\cal L}_{int} $ gives rise to the same physics as ${\cal L}_{int}'$ at the linearized level.

\section{Discussion and Conclusions}
We have analyzed scattering amplitudes in a model of emergent gravity with general field-space metric for scalar fields that play the role of clock and rulers after gauge-fixing. The classical equations of motion admit a background solution in which the emergent spacetime metric is equal to the field-space metric. The quantum theory then admits a perturbative expansion about this background, so that the theory describes an emergent quantum gravity about the prescribed spacetime background. In the case that the field-space metric is nearly flat, we demonstrated that scattering off of the background  spacetime is as in general relativity, as is 2-into-2 scattering through the exchange of a composite spin-2 graviton. 

We note that even if the regulator scale is taken to infinity (for example $\epsilon\rightarrow0$ in dimensional regularization), so that the effective $M_{{\rm Pl}}\rightarrow\infty$,  matter will still scatter off of the gauge-fixed clock and ruler fields in such a way that the field-space metric plays the role of the background spacetime.

It was important in our analysis that the dynamics of the clock and ruler fields was  due only to the field-space kinetic term. If the potential had depended on the fields $X^I$ then the classical backgrounds for the clock and rulers would generally not admit the
static-gauge condition $X^I\propto x^\mu\delta_\mu^I$. For example, oscillating configurations of a massive clock field would be bounded in magnitude and could not be transformed by a coordinate transformation to an unbounded solution like the static-gauge configuration.
The possibility of configurations that do not admit the static gauge condition also raises another issue. By assuming the static gauge we are only integrating over a subset of field configurations. These are configurations close to the classical background, so we suspect that these solutions dominate perturbative contributions to correlation functions. However, the contribution of other configurations, which are nonperturbative  in the present approach, deserve further investigation.

Finally, we note that because the linearized couplings of the matter fields to both the composite graviton state and to the background spacetime metric are through the energy-momentum tensor, an extension of the theory to include scalars with different masses is guaranteed to contain universal gravitational couplings. 

\begin{acknowledgments}  
This work was supported by the NSF under Grant PHY-1519644. We are grateful to Chris Carone and Diana Vaman for many useful conversations and for collaboration on related work. We also thank C.~Carone, D.~Vaman and Tangereen Claringbold for sharing their work that appeared in Ref.~\cite{Carone:2017mdw} around the same time as this work.
\end{acknowledgments}

\end{document}